\documentstyle[12pt,iopfts]{ioplppt}
\def\rf#1{(\ref{#1})}
\def\im{\mathop{{\rm Im}}}

\def\bbox#1{{\bi #1}}
\def\myrm{\vphantom{i}}

\newcommand{\stack}[2]
 {\stackrel{\scriptstyle #1}{#2}\hspace{-3.5pt}\vphantom{#2}}

\begin{document}

\title{ Quantization of fields over de Sitter space  \\
 by the method of generalized coherent states. \\
 I. Scalar field }[Quantization by the generalized coherent states. I]
\author{S A Pol'shin\ftnote{1}{E-mail: itl593@online.kharkov.ua}}
\address{Department of Physics, Kharkov State University, \\
  Svobody Sq., 4, 310077, Kharkov, Ukraine }
\jl{1}

\date{}
\begin{abstract}
A system of generalized coherent states for the de Sitter group obeying
Klein-Gordon equation and corresponding to the massive spin zero particles
over the de Sitter space is considered. This
 allows us to construct the quantized scalar field by the resolution over
these coherent states; the corresponding propagator can be computed by the
method of analytic continuation to the complexified de Sitter space and
coincides with expressions obtained previously by other methods. We show that
this propagator possess the de Sitter-invariance and causality properties.
\end{abstract}

\section{Introduction}

In the later years a considerable progress in the  theory of massive
scalar field over the de Sitter (dS) space  is attained due to
 using the new mathematical methods. In~\cite{24} is shown that the
two-point Wightman function ${\cal W}(x,y)$ which corresponds
to this field and obeys the conditions of causality,
dS-invariance and positive definiteness can be obtained as the
boundary value of the holomorphic function $W(z_1 ,z_2)$ defined over the
complexified dS space.  In turn, the function $W(z_1 ,z_2)$ can be
represented as an integral over so-called "plane waves" which obeys the
Klein-Gordon equation over the dS space and generalizes the usual plane waves
over the Minkowski space.  In~\cite{dS-PLB} we applied to examining the
quantum fields over the dS space the method of generalized coherent states
(CS) which has been fruitfully used in various physical problems
(see~\cite{coher1/4} and references therein). In the mentioned paper we shown
that the above "plane waves" are CS for the dS space to within  the
coordinate-independent multiplier, and their scalar product coincides with
the two-point function considered in~\cite{24}.

Nevertheless, some questions remain open. Can we construct the  quantized
field over the dS space by the expansion over the mentioned "plane waves" in
such a way that its propagator will be equal to ${\cal W}(x,y)-{\cal
W}(y,x)$? What is the explicit form of this propagator?
The purpose of the present paper is to answer on
these questions and explain systematically the results briefly mentioned
in~\cite{dS-PLB}.

The present paper is composed as follows. In section 2, bearing in mind the
application to the spinor field (see part~II of this series of papers),  we
give the method of construction of CS in the maximally general form for which
the Perelomov's definition is a special case. In section 3 we
consider the dS space, its symmetry group and the classification of its
irreducible representations. The realization of the dS group as a group of
transformations of Lemaitre coordinate system is given too.
Following~\cite{dS-PLB}, in section 4 we realize the dS group as a group of
transformations of  functions over ${\Bbb R}^3$ and then construct CS system
for the dS space which correspond to the massive spin zero particles and obey
the dS-invariant Klein-Gordon equation. The scalar product of two CS is the
two-point function considered in~\cite{24}; from here its dS-invariance
follows immediately which is proved in~\cite{24} by the other reasons.
The integral that defines this two-point function may be regularized passing
to the complexified dS space. For the sake of completeness we reproduce some
results of the J.Bros and U.Moschella's paper~\cite{24} and compute the
two-point function over the complexified dS space in the explicit form. In
section 5 we construct the quantized scalar field by the expansion over CS
constructed in section 4. The propagator of this field is the difference of
two-point function and the permuted one. We show that the boundary values on
the real dS space  of the two-point function computed in section 4 coincides
with the Green's function obtained previously by  other methods~\cite{76,33}.
The propagator which is the difference of two two-point functions coincides
with that  obtained previously starting from the demands of the dS-invariance
and the satisfaction of the Klein-Gordon equation and the boundary
conditions\cite{77/81/82}. Thus, the relation between  different
expressions for the propagator available in the literature is
established (review of the papers concerning the
propagators over dS space see in~\cite{caus4/5}).

\section{Definition of the CS system}

Let $\cal G$ is a Lie group and
${\cal G}\ni g\mapsto T(g)$ is its representation in a linear
vector space $H$ with operators $T(g)$. Consider some vector
$|\psi_0 \rangle\in H$ yielding the set of vectors
\[ \{ |\psi_g \rangle\equiv T(g)\psi_0,\ \forall g\in {\cal G} \}.\]
Define the equivalence relation $\sim$ between the vectors of the $H$
space coordinated with the  product over $H$ in the following way.
Let $|\xi'\rangle$ and $|\xi''\rangle$ are the vectors of $H$.
Then we assume the existence of a  product (which, in general, isn't
the mapping of $H\times H$ to ${\Bbb R}$) such as
\[\label{defsim}
|\xi'\rangle\sim|\xi''\rangle
\Rightarrow \langle\xi'|\xi'\rangle =\langle\xi''|\xi''\rangle.\]
Consider the subgroup $\cal H$ of $\cal G$ which remains in the
rest the equivalency class generated by $|\psi_0\rangle$:
\[ h\in {\cal H}\Longleftrightarrow T(h)|\psi_0\rangle \sim
|\psi_0\rangle .\]
It is obvious that the number of unequivalent elements of the mentioned set
$|\psi_g \rangle$ is less than the number of elements of the group $\cal G$
because  the elements $g$ and $gh,\ h\in {\cal H}$ generate the equivalent
vectors. Then, in fact, the set of unequivalent vectors is determined by the
set of all right equivalency classes $g\cal H$ which compose the symmetric
space ${\cal G}/{\cal H}$.

The mapping ${\cal G}/{\cal H}\ni \xi \mapsto g_\xi \in {\cal G}$
such that for an arbitrary $g_1 \in {\cal G}$ the equality
\begin{equation}\label{foliation}
g_1 g_\xi=g_{\xi '}h \qquad  h\in {\cal H}
\qquad  \xi ' =\xi_{g_1}
\end{equation}
is valid, is called the lifting from the
${\cal G}/{\cal H}$  space to the $\cal G$ group,
where $\xi\mapsto \xi_g$ is the action of $\cal G$ over the
${\cal G}/{\cal H}$ space. We shall use the following simple
method of construction of liftings. Let
$\xi_\circ$ is a "standard" point of the ${\cal G}/{\cal H}$ space.
Let us denote
as $g_\xi$ the set of transformations parametrized by  points $\xi$ of the
${\cal G}/{\cal H}$ space so as $\xi=(\xi_\circ )_{g_\xi}.$ It is easily seen
that
$\xi \mapsto g_\xi$ is a lifting. Indeed, let $g_1
\in {\cal G}$ is an arbitrary transformation from the group $\cal G$. Then
the transformations
$g_1 g_\xi$ and $g_{\xi'}$ both transform the point
$\xi_\circ$ into the point $\xi'$; then the transformation
$(g_1 g_{\xi})^{-1} g_{\xi'}$ remains the point $\xi_\circ$ in the rest and
 therefore belongs to $\cal H$.

The choice of lifting is the choice of the {\it representative}
$g_\xi \in\cal G$ for each equivalency class $\xi$. Then the set of all
 unequivalent vectors $|\psi_g\rangle$ is given by the {\it coherent states}
 system
\[ |\xi\rangle =T(g_\xi) |\psi_0\rangle .\]
The major property of CS system is its $\cal G$-invariance which follows
 from~(\ref{foliation}):
\begin{equation}\label{lor7}
T(g)|\xi\rangle \sim |\xi_g\rangle \qquad  g\in {\cal G}.
\end{equation}
The Perelomov's definition for the CS system is narrower than ours as
he suppose that $\sim$ is the equality to within the phase:
\[ |\xi'\rangle\sim|\xi''\rangle \Leftrightarrow |\xi'\rangle=\e^{\i\alpha}
|\xi''\rangle  \qquad \alpha\in {\Bbb R}.\]
In a certain sense, our definition is the further generalization of
a so-called vector-like CS~\cite{coher10}.  Another difference of our
definition from  Perelomov's one is that we, following~\cite{coher10}, do not
assume the compactness of the $\cal H$ subgroup.

\section{Representations of the de Sitter group}

The dS space is a four-dimensional hyperboloid determined by the equation
$\eta _{AB}x^{A}x^{B}=-R^{2}$
in the five-dimensional space
with the pseudo-euclidean metric
$\eta_{AB} \quad (A, B, \ldots=0\ldots 3, 5)$ of signature $(+ - - - -)$.
Except the explicitly covariant vierbein indices, all the ones are raised and
lowered by the Galilean metric tensors $\eta_{AB}$ and $\eta_{\mu \nu}$.
The metric in  coordinates $x^\mu$ has the form
\begin{equation}\label{5.  21}
 g_{\mu \nu}=\eta _{\mu \nu}-
\frac{x^{\mu}x^{\nu}}{R^{2}\chi ^{2}}   \qquad
g^{\mu \nu}=\eta ^{\mu \nu}+\frac{x^{\mu}x^{\nu}}{R^{2}}
\end{equation}
where $\chi  =(1+x\cdot x/R^{2})^{1/2}.$
The symmetry group of
the dS space is the dS group $SO(4,1)$ with ten generators
$J^{AB}=-J^{BA}$ obey commutation relations
\begin{equation}\label{5.  22}
[J_{AB}, J_{CD}]=\eta _{AD}J_{BC}+\eta _{BC}J_{AD}-
\eta _{AC}J_{BD}-\eta _{BD}J_{AC}.
\end{equation}
Let us denote $P^\mu =R^{-1}J^{5\mu}$; these generators correspond to
translations.

The action of an arbitrary element $g\in {SO(4,1)}$ of the dS group over the
dS space we denote as $x\mapsto x_g$. The stationary subgroup of an arbitrary
point of dS space is $SO(3,1)$; then we can identify the dS space with the
set of equivalency classes ${SO(4,1)}/{SO(3,1)}$.

Let us construct the operators
\begin{equation}\label{3.  2}
\Pi^\pm_i=P_{i}\pm\frac{1}{R}J_{0i}.
\end{equation}
Using the commutation relations~(\ref{5. 22}) it is easy to show that
\begin{equation}\label{3.  3}
[\Pi^+_i ,\Pi^+_k ] =[\Pi^-_i ,\Pi^-_k ] =0.
\end{equation}
We can take the operators $\bPi^{+},\bPi^{-},P^{0}$ and $J_{ik}$
as a new set of generators of the dS group; they generates subgroups which we
denote as ${ \cal T}^{+}, {\cal T}^{-},{\cal T}^{0}$ and ${\cal
R}=SO(3)$, respectively. The groups ${\cal T}^{\pm}$ are abelian by the virtue
of~(\ref{3.  3}). Besides~(\ref{3. 3}), the commutation relations are
\begin{equation} \label{3.  4}
\eqalign{
{[}\Pi^+_i ,\Pi^-_k {]}=
-\frac{2}{R}P^{0}\delta_{ik}+\frac{2}{R^2}J_{ik} \qquad
{[}P^{0},J_{ik}{]}=0 \\
{[}\Pi^\pm_i ,J_{kl}{]}=
\Pi^\pm_k \delta_{il}-\Pi^\pm_l \delta_{ik} \qquad
{[}P^{0},\bPi^{\pm} {]} =\pm \frac{1}{R}\bPi^{\pm}.
}
\end{equation}

The dS group has two independent Casimir operators:
 \begin{equation}\label{5.  26}\label{5.  27}
 C_{2}=-\frac{1}{2R^{2}}J_{AB}J^{AB} \qquad
C_{4}=W_{A}W^{A}
\end{equation}
where
\begin{equation}\label{5.  28}
 W_{A}=\frac{1}{8R}\varepsilon _{ABCDE}J^{BC}J^{DE}
\end{equation}
is an analog of the Pauly--Lubanski pseudovector in the Poincar\'e group.
There is two series of the dS group irreducible representations~\cite{4/5}:

 1)$ \bpi_{p, q}$, $p=1/2, 1, 3/2, \ldots
;q=p, p-1, \ldots, 1$ and
$1/2$.  The eigenvalues of Casimir operators in this series are
\begin{equation}\label{5.  29}
\eqalign{
R^{2}C_{2}=p(p+1)+q(q-1)-2 \\
\label{5.  30}
 R^{2}C_{4}=p(p+1)q(q-1).
}
 \end{equation}

 2)$\bnu_{m, s}$.
The quantity $s$ is a spin,
 $s=0, 1/2, 1, \ldots$;the quantity $m$ corresponds to a mass, at the
integer spin $m^{2}>0$; at the half-integer spin
$m^{2}>1/4R^{2}$; at $s=0 \quad m^{2}>-2/R^{2}$.
\begin{eqnarray}\label{5.  31}
 C_{2}=-m^{2}+R^{-2}(s(s+1)-2) \\
\label{5.  32}
C_{4}=-m^{2}s(s+1).
\end{eqnarray}

The generators of five-dimensional rotations are
\[ \label{5.  33}
J^{(l)AB}=(x^{A}\eta^{BC}-x^{B}\eta^{AC})\partial_{C}.\]
As the fifth coordinate is not independent:
$x^{5}=R \chi, $ then $\partial_{5}=0$ and we obtain the generators
of scalar representation:
\begin{equation}\label{5.  34}\label{3. 4a}
P^{(l)}_{\mu}=\chi \partial _{\mu} \qquad
J^{(l)\mu \nu}=(x^{\mu}\eta ^{\nu \sigma}-
x^{\nu}\eta ^{\mu \sigma})\partial _{\sigma}.
\end{equation}
They compose the representation $\bnu_{m,0}$ since
\begin{equation}\label{5.  35}
W_{A}^{(l)}=0 \Rightarrow  C_{4}^{(l)}=0.
\end{equation}
As $(-g)^{1/2}=1/\chi,$ then for the second order Casimir operator in the
scalar representation we obtain from
~\rf{5.  26} and~\rf{5.  34}:
\[ C_{2}^{(l)}=\Box \equiv  (-g)^{-1/2}\partial
_{\mu}((-g)^{1/2}g^{\mu \nu} \partial _{\nu}).\]
 Then using~(\ref{5. 31}) we obtain
that in the representation $\bnu_{m, 0}$ the Klein-Gordon equation
\begin{equation}\label{5.  36} \label{5. 37}
(\Box +m^{2}+2R^{-2})\psi  =0
\end{equation}
is satisfied.

By the virtue of~(\ref{3.  3}) in the scalar representation the
generators~(\ref{3.  2}) are derivatives along certain new coordinates
called the Lemaitre coordinates:
\begin{equation}\label{3.  4b}
\bPi^\pm=\frac{\partial}{\partial {\bi y}_\pm}.
\end{equation}
Substituting~(\ref{3. 4a}) and~(\ref{3. 4b}) into r.h.s. and l.h.s
of equation (\ref{3. 2}) respectively, we obtain the connection of
${\bi y}_\pm$ with $x^\mu$. A new time coordinate independent on
${\bi y}_{\pm}$ is denoted as
$y^{0}_{\pm}=\tau_{\pm}$; then the transformation rules from the old
coordinates to the new ones are
\begin{equation}\label{3.  6a}
{\bi y}_{\pm}={\bi x}\e^{\mp\tau_{\pm}/R} \qquad
\e^{\pm\tau_{\pm}/R}=\chi \pm \frac{x^0}{R}.
\end{equation}
The operator $P_{0}$   in  new coordinates takes the form
\[ P_{0}^{(l)}=\frac{\partial}{\partial \tau_{\pm}}\mp
\frac{1}{R}{\bi y}_{\pm}\frac{\partial}{\partial {\bi y}_{\pm}}.\]
The finite transformations from the groups
${\cal T}^{\pm}$ and ${\cal T}^{0}$,
which we denote as $\Theta_{\pm}$ and $\Theta_{0}$, act in the scalar
representation in the following way:
\begin{equation}\label{3.  9}
\label{3.  10}
\eqalign{
 g= \Theta_{\pm}({\bi a})\equiv \exp (\bPi^{\pm}{\bi a}R) \ :
\quad \left\{
\begin{array}{l}
{{\bi y}}_{\pm}\longmapsto {{{\bi y}}'}_{\pm}={{\bi y}}_{\pm}+{{\bi a}}R \\
\tau_{\pm}\longmapsto {\tau'}_{\pm}=\tau_{\pm}
\end{array}   \right. \\
 g= \Theta_{0}(\varepsilon )\equiv \exp (P_0 \varepsilon R)\ :
\quad\left\{
\begin{array}{l}
{{{\bi y}}'}_{\pm}={{\bi y}}_{\pm}\e^{\mp \varepsilon} \\
{\tau'}_{\pm}=\tau_{\pm}+\varepsilon R.
\end{array}  \right.
}
\end{equation}
We assume that the transformations act in the order from the right to the
left.

\section{Scalar coherent states}
The dS group is isomorphic to the group of conformal transformations of the
three-dimensional real space.
We denote the vector of this space  as ${\bi w}$. There exist two different
conformal realizations of the dS group; the first one corresponds
to the upper, and the second one to the lower signs in the following
formulas. The generators take the form
\begin{eqnarray}
R\bPi^{\mp}=-\frac{\partial}{\partial {{\bi w}}} \qquad
R\bPi^{\pm}={w}^{2}\frac{\partial}{\partial {{\bi w}}}-2{{\bi w}}
({{\bi w}}\frac{\partial}{\partial {{\bi w}}}) \nonumber\\
R P_{0}=\pm {{\bi w}}\frac{\partial}{\partial {{\bi w}}} \qquad
J_{ik}={w}_{k}\frac{\partial}{\partial  {w}_{i}}-
{w}_{i}\frac{\partial}{\partial {w}_{k}}. \nonumber
\end{eqnarray}
They obey commutation relations~(\ref{3.  3}) and~(\ref{3.  4}).
 Finite transformations have the form
\begin{eqnarray}
 g=\Theta_{\mp}({{\bi a}}):\ {{\bi w}}_{g}={{\bi w}}-{{\bi a}}\nonumber \\
\label{3.  15}
g=\Theta_{\pm}({{\bi a}}):\ {{\bi w}}_{g}=
\frac{{{\bi w}}+{{\bi a}}{w}^{2}}{1+2{\bi wa}+{w}^{2}{a}^{2}} \\
g=\Theta_{0}(\varepsilon ):\ {{\bi w}}_{g}={{\bi w}}\e^{\pm \varepsilon}
\nonumber
\end{eqnarray}
where we used the notations~(\ref{3. 9}) for finite
transformations of the dS group.

Let us define two different representations of the dS group acting over the
space of functions dependent on ${\bi w}$:
\[\label{3.  20}
T^{\pm}_\sigma (g)f({{\bi w}})=\left(\alpha^{\pm}_{{\bi w}}(g)\right)^\sigma
f({{\bi w}}_{g^{-1}})\]
where $\sigma\in {\Bbb C}$ and
\[\alpha^{\pm}_{{\bi w}}(g)=
\det \left( \frac{\partial {\rm w}^{i}_{g^{-1}}}{\partial {\rm w}^{k}}
\right)^{-1/3}=
\left\{
\begin{array}{ll}
1 & g\in {\cal T}^{\mp}\circledS {\cal R} \\
\e^{\pm \varepsilon} & g=\Theta_{0}(\varepsilon) \\
1-2{\bi aw}+a^{2} w^{2} & g=\Theta_{\pm}({{\bi a}})
\end{array}
\right.\]
and the action of the dS group over the
${\Bbb R}^3$ space is determined by~(\ref{3.  15}).
We denote these representations as $T_\sigma^\pm$. It is easily seen that the
generators in these representations are
\begin{equation}\label{3. 20a}
\eqalign{
R\bPi^{\mp}=-\frac{\partial}{\partial {{\bi w}}} \qquad
R\bPi^{\pm}={w}^{2}\frac{\partial}{\partial {{\bi w}}}-2{{\bi w}}
({{\bi w}}\frac{\partial}{\partial {{\bi w}}})+2\sigma {{\bi w}} \\
R P_{0}=\pm\left( {{\bi w}}\frac{\partial}{\partial {{\bi w}}}-\sigma \right)
\qquad
J_{ik}={w}_{k}\frac{\partial}{\partial  {w}_{i}}-
{w}_{i}\frac{\partial}{\partial {w}_{k}}.
}
\end{equation}
We define the scalar product in the space of  representation
$T_\sigma^\pm$ as follows:
\[\langle f_{1} |f_{2}\rangle =\int_{{\Bbb R}^3} \d^{3}{{\bi w}} \,
f_{1}^{*}({{\bi w}}) f_{2}({{\bi w}}).\]
It is not difficult to show that it is dS-invariant at
\[\sigma=\sigma_{0}\equiv -\frac{3}{2}-\i\mu R \qquad  \mu\in {\Bbb R}.\]
Then the representation $T^\pm_{\sigma_0}$ is unitary; but it is reducible
since we do not assume the square integrability of functions carrying
it and therefore the space contains an invariant subspace of square integrable
functions. Such an extension of the representation space is necessary
for the construction of CS with noncompact stability groups~\cite{coher10}.

The equality
\[\label{3. 18a}
g_{y_{\pm}}=\Theta_{\pm}({{\bi y}}_{\pm}/R)
\Theta_{0}(\tau_{\pm}/R)\]
defines a lifting in the sense of~(\ref{foliation}) since the transformation
$g_{y_\pm}$ transforms the origin into the point with coordinates $y_\pm$.
As an equivalency relation we can take the equality. Then the vector
$|\psi_{0}\rangle$ being Lorentz-invariant under the action of the
representation $T^\pm_\sigma$ is $|\psi_{0}\rangle
=(1-{w}^{2})^{\sigma}.$ Then we can construct the CS system
\[\label{3. 24}
|x ,\pm;\sigma\rangle =T^{\pm}_\sigma (g_{y_{\pm}(x)})|\psi_{0}\rangle \]
where we assume that  the Lemaitre coordinates are dependent on $x^\mu$
by the transformations~(\ref{3.  6a}).  The explicit form of CS as functions
of ${\bi w}$ is
\[|x ,\pm;\sigma\rangle \equiv \Phi^{(0)\pm}_{{\bi w}}(x;\sigma) =
(1-{w}^{2})^{\sigma}\varphi_{k_{{\bi w}}}^{(0)\pm}(x;\sigma)\]
where
\[\varphi_{k}^{(0)\pm}(x;\sigma)=
\left( \chi\pm \frac{k\cdot x}{R} \right)^{\sigma}\label{5. 58} \qquad
k^{\mu}_{{\bi w}}=\left( \frac{1+{w}^{2}}{1-{w}^{2}},
 \pm \frac{2{{\bi w}}}{1-{w}^{2}}\right) \]
then $k_{{\bi w}}\cdot k_{{\bi w}}=1$.
From~(\ref{lor7}) the transformation rules
\begin{equation}\label{transf}
\Phi_{{\bi w}}^{(0)\pm}(x_g ;\sigma)=\left(
\alpha^\pm_{{{\bi w}}}(g)\right)^\sigma \Phi_{{{\bi w}}'}^{(0)\pm}(x;\sigma)
\qquad  {{\bi w}}'={{\bi w}}_{g^{-1}}
\end{equation}
follow.

Functions $\varphi_{k}^{(0)\pm}(x;\sigma_0)$ obey the dS-invariant
Klein-Gordon equation~(\ref{5. 36}) and were known previously in this
capacity~\cite{24,coher9/78}.  Under $R \rightarrow \infty$ these functions
pass into the usual plane waves over the Minkowski space.

Let us consider the scalar product of two CS in the representation
$T^\pm_{\sigma_0}$; it is easily seen that the scalar products in
representations $T^+_{\sigma_0}$ and
$T^-_{\sigma_0}$ are equal to each other. This may be proved
considering the inversion $\bbox{w}\mapsto
-\bbox{w}/{\myrm w}^2$ with which
\[\Phi^{(0)\pm}_{\bbox{w}}(x;\sigma_0)\mapsto (-{\myrm w}^2)^{-\sigma_0}
\Phi^{(0)\mp}_{\bbox{w}}(x;\sigma_0).\]
Then a two-point function can be defined as
\[\label{twop-0-def}
\langle \stack{2}{x},\pm;\sigma_0|\stack{1}{x} ,\pm;\sigma_0 \rangle =
\int_{{\Bbb R}^3}\d^3 {{\bi w}}\, \Phi_{{\bi w}}^{(0)\pm}(\stack{1}{x};\sigma_0 )
\Phi_{{\bi w}}^{(0)\pm}(\stack{2}{x};\sigma_0^* )= \frac{1}{8}
{\cal W}^{(0)}(\stack{1}{x},\stack{2}{x}).\]
It is dS-invariant by the virtue of unitarity of the representation
$T^\pm_{\sigma_0}$:
\[{\cal W}^{(0)}
(\stack{1}{x}_{g},\stack{2}{x}_{g})={\cal W}^{(0)}(\stack{1}{x},\stack{2}{x})
\qquad  g\in SO(4,1).\]
We can obtain an another expression for
${\cal W}^{(0)}(\stack{1}{x},\stack{2}{x})$ passing to the integration over
3-sphere~\cite{24}. Let us consider a unit euclidean four-vector
$l_a$, $a,b=1,2,3,5$ dependent on the three-vector ${\bi w}$:
\[l^a_{{\bi w}}=\left( \mp \frac{2{{\bi w}}}{1+{\rm w}^2},
\frac{1-{\rm w}^2}{1+{\rm w}^2}\right) \qquad
l^a_{{\bi w}}l^a_{{\bi w}}=1. \]
Then computing the Jacobian of the transformation from
${\bi w}$ to ${\bi l}_{{\bi w}}$ we obtain
\begin{equation}\label{twop-eucl}
{\cal W}^{(0)}(\stack{1}{x},\stack{2}{x})=\int_{S^3}\frac{\d^3{\bi l}}{l^5}
\left( \frac{\stack{1}{x}^0+l^a\stack{1}{x}^a}{R}
\right)^{-\i\mu R-3/2}
\left( \frac{\stack{2}{x}^0+l^a\stack{2}{x}^a}{R}
\right)^{\i\mu R-3/2}.
\end{equation}
The function ${\cal W}^{(0)}(\stack{1}{x},\stack{2}{x})$
coincides with the two-point function over the dS space considered
in~\cite{24}.  In general, the determining integrals diverge because of
$|\psi_0\rangle$ is not square-integrable over ${\Bbb R}^3$. We can make the
integral meaningful by passing to the complexified
dS space with subsequent computation of the boundary values over the real dS
space~\cite{24}. In this case the dS-invariance of two-point function remains
valid since the transformation rules~(\ref{transf}) remains correct.

Let us consider the domain ${\cal D}^\pm$ in the complexified dS space
(we shall denote its points as $\zeta$) defined as
\[ \quad \pm\im \zeta^0 >0 \qquad
\im \zeta^A \im \zeta_A >0.\]
The domain ${\cal D}^+$ (${\cal D}^-$) is the domain of analyticity of
functions
$\varphi^{(0)\pm}_{k}(\zeta;\sigma_0)$
($\varphi^{(0)\pm}_{k}(\zeta;\sigma_0^*)$). Then the integral
(\ref{twop-eucl}) converges at $\stack{1}{\zeta}\in {\cal D}^+$ and
$\stack{2}{\zeta}\in {\cal D}^-$ since 3-sphere volume is finite.
Let us choose the points as
\begin{equation}\label{points-dS}
\stack{1}{\zeta}^A=(\i\cosh v,{\bi 0},i\sinh v) \qquad
\stack{2}{\zeta}^A=(-\i,{\bi 0},0) \qquad  v\in {\Bbb R}.
\end{equation}
Then using the formula~\cite{59}
\begin{equation}\label{BE123-8}
\mathop{_2 F_1} (a,b;c;z)=
\frac{2^{1-c}\Gamma(c)}{\Gamma(b)\Gamma(c-b)}
 \int_{0}^\pi \d\varphi
\, \frac{(\sin\varphi)^{2b-1}(1+\cos\varphi)^{c-2b}}{\left(
1-\frac{z}{2}+\frac{z}{2}\cos\varphi\right)^a}
\end{equation}
we obtain
\[\label{twop-cdS}
{\cal W}^{(0)}(\stack{1}{\zeta},\stack{2}{\zeta})=
-\frac{\pi^2}{2} \e^{-\pi\mu R}\mathop{_2 F_1}\left( -\sigma_0^*,
-\sigma_0 ;2;\frac{1-\rho}{2}\right)\]
where $\rho=R^{-2} \stack{1}{\zeta}^A \stack{2}{\zeta}_A$.
The expression obtained in~\cite{24} is in fact equivalent to
the above expression
to within a constant.

\section{Spin zero field over de Sitter space\label{field0-dS}}

Let us define the quantized spin zero field in the dS space as
\[\phi^{(0)}(x)=\int_{{\Bbb R}^3} \d^3 {{\bi w}}\, \left(
\Phi^{(0)+}_{{\bi w}}(x;\sigma_0 )a^{(+)} ({{\bi w}})+
\Phi^{(0)-}_{{\bi w}}(x;\sigma_0^* )a^{(-)\dagger}
({{\bi w}})\right)\]
where  $a^{(\pm)}({{\bi w}})$ and $a^{(\pm)\dagger} ({{\bi w}})$ are two sets of
bosonic creation-annihilation operators with the commutation relations
\[\label{comm-bosonic}
[a^{(\pm)}({{\bi w}}), a^{(\pm)\dagger} ({{\bi w}}')]=
\delta ({{\bi w}},{{\bi w}}') \]
where $\delta({\bbox{w}}_1 ,{\bbox{w}}_2)$ is the
$\delta$-function over ${\Bbb R}^3$.
Now  compute the propagator
\begin{equation}\label{3. 31a}
\Bigl[ \phi^{(0)}(\stack{1}{x}),\phi^{(0)\dagger}(\stack{2}{x})\Bigr] \equiv
\frac{1}{8}G^{(0)}(\stack{1}{x},\stack{2}{x})=
\frac{1}{8}\left( {\cal W}^{(0)}(\stack{1}{x},\stack{2}{x})-
{\cal W}^{(0)}(\stack{2}{x},\stack{1}{x}) \right)
\end{equation}
in the explicit form by passing to the complexified dS space.
Considering the real dS space as the boundary of the domains
$(\stack{1}{\zeta},\stack{2}{\zeta})\in {\cal D}^\pm \times {\cal D}^\mp$,
let us denote the  boundary values of the functions
${\cal W}^{(0)}(\stack{1}{\zeta},\stack{2}{\zeta})$
as ${\cal W}^{(0)\pm}(\stack{1}{x},\stack{2}{x})$. To compute these functions
we put
$\stack{1}{\zeta}=\stack{1}{x}+\i\stack{1}{\epsilon}$ and
$\stack{2}{\zeta}=\stack{1}{x}-\i\stack{2}{\epsilon},$
where $\stack{1}{\epsilon}$ and $\stack{2}{\epsilon}$ are two real
infinitesimal time-like forward four-vectors and then indeed
$(\stack{1}{\zeta},\stack{2}{\zeta})\in {\cal D}^+ \times {\cal D}^-$.
It is easily seen that
\[\stack{1}{\zeta}^A \stack{2}{\zeta}_A =\stack{1}{x}^A \stack{2}{x}_A +
\frac{\i}{\stack{1}{x}^5 \stack{2}{x}^5}(\stack{1}{\epsilon}+
\stack{2}{\epsilon})\cdot \left( \frac{\stack{2}{x}}{\stack{2}{x}^5}-
\frac{\stack{1}{x}}{\stack{1}{x}^5}\right).\]
Then under the above assumptions the sign of the imaginary part of
$\stack{1}{\zeta}^A \stack{2}{\zeta}_A$ does
not depend on the way in which
$\stack{1}{\epsilon}$ and $\stack{2}{\epsilon}$ tend to zero.
Let $\stack{2}{x}^\mu =0$ and $\stack{1}{x}\cdot \stack{1}{x}\geq 0$, then
\[\stack{1}{\zeta}^A \stack{2}{\zeta}_A =\stack{1}{x}^A \stack{2}{x}_A -
i0\varepsilon (\stack{1}{x}^0).\]
The case of backward
$\stack{1}{\epsilon}$ and $\stack{2}{\epsilon}$ (then
$(\stack{1}{\zeta},\stack{2}{\zeta})\in {\cal D}^- \times {\cal D}^+$)
may be considered in the completely analogous way. Then
\[{\cal W}^{(0)\pm}(\stack{1}{x},\stack{2}{x})=
-\frac{\pi^2}{2} \e^{-\pi\mu R}\mathop{_2 F_1}\left( -\sigma_0^*,
-\sigma_0 ;2;\frac{1-G \pm \i 0 \varepsilon(\stack{1}{x}^0)}{2}\right)\]
where $G=R^{-2}\stack{1}{x}^A \stack{2}{x}_A$. As
$(\stack{1}{\zeta},\stack{2}{\zeta})\in {\cal D}^+ \times {\cal D}^-$ yields
$(\stack{2}{\zeta},\stack{1}{\zeta})\in {\cal D}^- \times {\cal D}^+$,
then we get
\begin{equation}\label{x1x2-x2x1}
{\cal W}^{(0)+}(\stack{1}{x},\stack{2}{x}) =
{\cal W}^{(0)-}(\stack{2}{x},\stack{1}{x})
\end{equation}
(cf. proposition 2.2 of~\cite{24}) and by the virtue of~(\ref{3. 31a})
the propagator is equal to
\[ G^{(0)}(\stack{1}{x},\stack{2}{x})=
{\cal W}^{(0)+}(\stack{1}{x},\stack{2}{x})-
{\cal W}^{(0)-}(\stack{1}{x},\stack{2}{x}).\]
The function
${\cal W}^{(0)-}(\stack{1}{x},\stack{2}{x})$ coincides to within the constant
with the propagator obtained in~\cite{76} from the demands of satisfaction of
the Klein-Gordon equation and the boundary conditions. This function may be
obtained by the summation over the modes~\cite{33} and by the method of
discrete lattice~\cite{85} too.

As we assume that $\stack{1}{x}\cdot \stack{1}{x}\geq 0$ then
$\frac{1-G}{2}\geq 1$, but the integral~(\ref{BE123-8}) with which we define
the hypergeometric function, makes no sense at
$z\in [1,+\infty)$ and then  demands the analytic continuation
in the domain which contains the mentioned interval. To this end we shall use
the formulas~\cite{59}
\begin{eqnarray}
u_1 =\frac{\Gamma (c)\Gamma(b-a)}{\Gamma(c-a)\Gamma(b)}u_3 +
\frac{\Gamma(c)\Gamma(a-b)}{\Gamma(c-b)\Gamma(a)} u_4 \label{u1}\nonumber \\
u_2 =\frac{\Gamma (a+b+1-c)\Gamma(b-a)}{\Gamma(b+1-c)\Gamma(b)}
\e^{\mp \i\pi a}u_3 +
\frac{\Gamma(a+b+1-c)\Gamma(a-b)}{\Gamma(a+1-c)\Gamma(a)}
\e^{\mp \i\pi b} u_4  \nonumber
\end{eqnarray}
where the upper or lower sign should be chosen depending on the $\im z$
greater or smaller  than zero, and $u_1,\ldots,u_4$ are the Kummer solutions
of the hypergeometric equation:
\begin{eqnarray}
u_1 =\mathop{_2 F_1}(a,b;c;z)\nonumber \\
u_2 =\mathop{_2 F_1}(a,b;a+b+1-c;1-z)\nonumber \\
u_3 =(-z)^{-a}\mathop{_2 F_1}(a,a+1-c;a+1-b;z^{-1})\nonumber \\
u_4 =(-z)^{-b}\mathop{_2 F_1}(b,b+1-c;b+1-a;z^{-1}) .\nonumber
\end{eqnarray}
The functions $u_1 ,u_3, u_4$ are holomorphic at $z<0$. Then
at $a+b+1=2c$
\begin{equation}
\left. u_2 \right|_{z-i0}^{z+i0} =\i (\e^{\pi\mu R}+\e^{-\pi\mu R})\theta (-z)
u_1 \qquad  z\not= 0
\end{equation}
and thus we obtain for $G\not= -1$
\[ \fl
G^{(0)}(\stack{1}{x},\stack{2}{x})=-\frac{\i\pi^2}{2}(1+\e^{-2\pi\mu R})
\varepsilon (\stack{1}{x}^0 -\stack{2}{x}^0)
\theta \left(-\frac{1+G}{2}\right) \mathop{_2 F_1}
\left(-\sigma_0 ,-\sigma_0^* ;2; \frac{1+G}{2}\right). \]
To within the constant, the above expression coincide with that obtained
previously from the demands of satisfaction of the
Klein-Gordon equation and the boundary conditions~\cite{77/81/82}. The only
difference is that our method do not allow us to find the behavior of
the propagator over the "light cone" $G=-1$.

\ack

I am grateful to Yu P Stepanovsky for the constant support during
the fork and to W Drechsler and Ph Spindel sending me  copies of their
papers~\cite{coher9/78,caus4/5}.

\end{document}